\documentstyle[prl,aps,twocolumn,floats,epsf]{revtex}

\newcommand{\npb}[2]{{\em Nucl. Phys.}              {\bf B#1}, #2 }

\newcommand{\prt}[2]{{\em Phys. Rev.}               {\bf D#1}, #2 }
\newcommand{\pru}[2]{{\em Phys. Rev. Lett.}         {\bf  #1}, #2 }

\newcommand{\rmq}[2]{{\em Rev. Mod. Phys.}          {\bf  #1}, #2 }
\newcommand{\etal}{{\em et al.}}

\newcommand{\ovl}{\overline}
\newcommand{\be}{\begin{equation}}
\newcommand{\ee}{\end{equation}}
\newcommand{\ba}{\begin{array}}
\newcommand{\ea}{\end{array}}

\def\ms{\mbox{{\footnotesize{$\overline{\rm MS}$ }}}}

\begin{document}

\preprint{
\noindent
\hfill
\begin{minipage}[t]{3in}
\begin{flushright}
IF--UNAM-- \\
\vspace*{2cm}
\end{flushright}
\end{minipage}
}

\draft

\title{Electroweak Radiative Corrections to Semileptonic $\tau$ Decays}

\author{Jens Erler}

\address{Instituto de F{\'\i}sica, Universidad Nacional Aut\'onoma de M\'exico,
04510 M\'exico D.F., M\'EXICO \vspace{1pt} \\ 
{\rm e-mail: erler@fisica.unam.mx} \vspace{1pt} } 

\date{November 2002}

\maketitle

\begin{abstract}
I present an update on the electroweak radiative correction factor to 
semileptonic $\tau$ decays, including a next-to-leading order resummation of 
large logarithms.  My result differs both qualitatively and quantitatively
from the one recently obtained by Davier~\etal.  As two consequences, (i) the 
discrepancy between the predictions for the muon $g-2$ based on $\tau$ decay 
data and $e^+ e^-$ annihilation data increases, and (ii) the $g-2$ prediction 
based on $\tau$ decay data appears to be consistent (within about one standard 
deviation) with the experimental result from BNL. 
\end{abstract}


The largest theoretical uncertainty in the Standard Model prediction of
the anomalous magnetic moment of the muon, $a_\mu = (g_\mu - 2)/2$, arises from
the hadronic two-loop vacuum polarization contribution, 
$\Delta a_\mu^{\rm had,(2)}$.  This contribution is two orders of magnitude 
larger than the ultimate experimental error anticipated by the Muon $g-2$ 
Collaboration at BNL~\cite{Bennett:2002jb}, so it needs to be controlled at 
the 1\% level or better. While non-perturbative QCD effects prevent a first 
principles calculation, $\Delta a_\mu^{\rm had,(2)}$ can be rigorously obtained
experimentally from a dispersion relation which relates it to an integral over
$e^+ e^-$ annihilation cross sections. Using the conserved vector current (CVC)
hypothesis one can obtain additional information by studying the invariant mass
distribution of $\tau$ decay hadronic final states.  This necessitates 
a careful assessment of CVC breaking effects, which was done in a recent 
article by Davier~\etal~\cite{Davier:2002dy}.  In this note, I present 
an update of the short distance electroweak radiative corrections to $\tau$ 
decays, representing a particular CVC breaking effect. This update is motivated
by two mistakes in one of the formulas of Ref.~\cite{Davier:2002dy}. 
Numerically, the corresponding shifts are modest, but not negligible, and have 
the same sign.

The leading electroweak radiative corrections to $\tau$ decays are enhanced by 
a large logarithm~\cite{Sirlin:1977sv,Marciano:vm},
\be 
   S_{\rm EW} = 
   1 + {3 \alpha\over 4 \pi} (1 + 2\ovl{Q}) \ln {M_Z^2\over m_\tau^2} = 1.01878
\label{sew1}
\ee
where $M_Z = 91.1876(2)$~GeV~\cite{Abbaneo:2001ix} is the $Z$ boson mass, and 
$\alpha = \alpha(m_\tau) = 1/133.50(2)$~\cite{Erler:1998sy} is the QED coupling
at the $\tau$ lepton mass, $m_\tau = 1776.99(3)$~MeV~\cite{Hagiwara:pw},
evaluated in the \ms renormalization scheme.  $\ovl{Q}$ is 
the hypercharge of the weak doublet produced in the final state. Therefore, 
$\ovl{Q} = 1/6$ for semileptonic decays, 
$\tau^- \rightarrow \nu_\tau \bar{u} d (s)$. Since $\ovl{Q} = - 1/2$ for
leptons, there are no large logarithms for leptonic $\tau$ decays.

The remaining (not logarithmically enhanced) corrections at ${\cal O} (\alpha)$
have been obtained in Ref.~\cite{Braaten:1990ef} (final state fermion masses 
are neglected throughout).  In the notation of Eq.~(17) of 
Ref.~\cite{Davier:2002dy} they are,
\begin{eqnarray}
   S_{\rm EW}^{\rm sub,had} = 1 + 
   {\alpha (m_\tau)\over \pi} \left({85\over 24} - {\pi^2\over 2} \right), \\ 
   S_{\rm EW}^{\rm sub,lep} = 1 +
   {\alpha (m_\tau)\over \pi} \left({25\over 8}  - {\pi^2\over 2} \right),
\label{sew2a}
\end{eqnarray}
for semileptonic and leptonic decays, respectively. 
In Ref.~\cite{Davier:2002dy}, however, $S_{\rm EW}^{\rm sub,had}$ was 
erroneously identified with the ratio,
\be
   {S_{\rm EW}^{\rm sub,had} \over S_{\rm EW}^{\rm sub,lep}} = 1 + 
   {5\over 12}{\alpha (m_\tau)\over \pi} = 1.00099.
\label{sew2b}
\ee
This amounts to a double counting
of the correction $S_{\rm EW}^{\rm sub,lep} - 1 = - 0.00432$: the hadronic 
spectral functions are normalized relative to the leptonic branching ratio (see
Eq.~(10) of Ref.~\cite{Davier:2002dy}) so that the ratio~(\ref{sew2b}) must be
included, but it is incorrect to perform an additional division by 
$S_{\rm EW}^{\rm sub,lep}$.

Since numerically, 
\be
  {\alpha_s\over \pi} \ln {M_Z^2\over m_\tau^2} \sim {\cal O}(1),
\label{aasln}
\ee
short distance QCD effects are of similar size as the ${\cal O} (\alpha)$ 
corrections discussed in the previous paragraph. They have been computed in 
Ref.~\cite{Sirlin:1981ie} and modify Eq.~(\ref{sew1}),
\be 
   S_{\rm EW} = 1 + {3 \alpha\over 4 \pi} \ln {M_Z^2\over m_\tau^2} 
   \left[ (1 + 2\ovl{Q}) - 2 \ovl{Q} {\alpha_s\over \pi} \right].
\label{sew3}
\ee
The short distance QCD correction corresponding to the term proportional to 
the strong coupling constant, $\alpha_s$, has been 
approximated~\cite{Sirlin:1981ie} in order to obtain an analytic result. 
I have checked that this approximation reproduces the exact 
${\cal O} (\alpha\alpha_s \ln M_Z^2)$ result within about 1\%. 
Since two scales enter Eq.~(\ref{sew3}), it is 
clear that a next-to-leading order renormalization group analysis is in order.

Resummation of the leading order logarithms in Eq.~(\ref{sew1}) is done using 
the renormalization group equation (RGE)~\cite{Marciano:pd},
\be
   \left[ \mu^2 {\partial\over\partial\mu^2} + \beta_0^{(1)} {\alpha^2\over\pi}
   {\partial\over\partial\alpha} - {\alpha\over\pi} \right] S(\mu_0,\mu) = 0,
\label{RGE}
\ee
where $\beta_0^{(1)}$ is the lowest order QED $\beta$-function coefficient. 
This RGE is subject to the initial condition, $S(\mu_0,\mu_0) = 1$, and its 
solution is given by,
\be
   S(\mu_0,\mu) = \left[ 1 - {\alpha(\mu_0)\over\pi} \beta_0^{(1)}
   \ln {\mu^2\over\mu_0^2} \right]^{ - {1\over\beta_0^{(1)}}}.
\ee
Applied to the case at hand, this is often written as,
\be
\ba{rcl}
   S(m_\tau,M_Z) &=& \left[{\alpha(m_b)\over\alpha(m_\tau)}\right]^{ 9\over 19}
                     \left[{\alpha(M_W)\over\alpha(m_b)}   \right]^{ 9\over 20}
                     \left[{\alpha(M_Z)\over\alpha(M_W)}   \right]^{36\over 17}
\vspace*{8pt} \\ &=& 1.01937,
\label{sew4}
\ea
\ee
where the solution to the one-loop RGE of QED,
\be
   \mu^2 {{\rm d}\over {\rm d}\mu^2} \alpha(\mu) = 
   \beta_0^{(1)} {\alpha^2(\mu)\over\pi},
\label{RGEalpha}
\ee
has been employed. It should be stressed, that consistency with the RGE demands
one-loop evolution of $\alpha (\mu)$ {\em within\/} each of the factors in 
Eq.~(\ref{sew4}). On the other hand, the values used {\em across\/} the various
factors, may be related to each other either by one-loop evolution or including
higher order running effects, since the difference is of higher order in 
the RGE~(\ref{RGE}). The increase of 
$S(m_\tau,M_Z)$ in Eq.~(\ref{sew4}) relative to Eq.~(\ref{sew1}) due to 
the summation of ${\cal O} (\alpha^n \ln^n M_Z^2)$ effects is about 3\% of 
the non-resummed correction.  

I will now extend the RGE analysis of the previous paragraph to properly sum up
all logarithms of ${\cal O} (\alpha\alpha_s^n \ln^n M_Z^2)$. Eq.~(\ref{RGE}) is
to be replaced by,
\be
   \left[ \mu^2 {{\rm d}\over {\rm d}\mu^2} - {\alpha\over\pi} 
   \left( 1 - {\alpha_s\over 4\pi} \right) \right] S(\mu_0,\mu) =
\label{RGENLO}
\ee
$$
     \left[                       \mu^2 {\partial\over\partial\mu^2} 
   + \beta_0^{(1)} {\alpha^2  \over\pi} {\partial\over\partial\alpha} 
   - \beta_0^{(3)} {\alpha_s^2\over\pi} {\partial\over\partial\alpha_s}
   - {\alpha\over\pi} \left( 1 - {\alpha_s\over 4\pi} \right) \right] S = 0,
$$
where $\beta_0^{(3)}$ is the lowest order QCD $\beta$-function coefficient.
With the definitions,
\be
\ba{rcl}
   \eta_\tau &=& {\alpha_s(m_\tau)\over 4\pi} {\left[ 1 + {75\over 76} 
   {\alpha_s(m_\tau)\over\alpha(m_\tau)} \right]}^{-1},
\vspace*{8pt} \\
   \eta_b    &=& {\alpha_s(m_b)   \over 4\pi} {\left[ 1 + {69\over 80} 
   {\alpha_s(m_b)   \over\alpha(m_b)}    \right]}^{-1},
\vspace*{8pt} \\
   \eta_W    &=& {\alpha_s(M_W)   \over 4\pi} {\left[ 1 + {69\over 17} 
   {\alpha_s(M_W)   \over\alpha(M_W)}    \right]}^{-1},
\ea
\ee
Eq.~(\ref{RGENLO}) is solved by,
\be
\ba{rcl}
   S(m_\tau,M_Z) &=& 
   \left[{\alpha(m_b)\over\alpha(m_\tau)}\right]^{{ 9\over 19}(1-\eta_\tau)}
   \left[{\alpha_s(m_b)\over\alpha_s(m_\tau)}\right]^{{ 9\over 19}\eta_\tau}
\vspace*{8pt} \\ & &
   \left[{\alpha(M_W)\over\alpha(m_b)}   \right]^{{ 9\over 20}(1-\eta_b)}
   \left[{\alpha_s(M_W)\over\alpha_s(m_b)}   \right]^{{ 9\over 20}\eta_b}
\vspace*{8pt} \\ & &
   \left[{\alpha(M_Z)\over\alpha(M_W)}   \right]^{{36\over 17}(1-\eta_W)}
   \left[{\alpha_s(M_Z)\over\alpha_s(M_W)}   \right]^{{36\over 17}\eta_W}
\vspace*{8pt} \\ &=& 1.01907 \pm 0.00001,
\label{sew5}
\ea
\ee
where I used the solutions to the one-loop RGE of QCD,
\be
   \mu^2 {{\rm d}\over {\rm d}\mu^2} \alpha_s(\mu) = 
   - \beta_0^{(3)} {\alpha_s^2(\mu)\over\pi},
\label{RGEalphas}
\ee
and of QED\footnote{QCD corrections to Eq.~(\ref{RGEalpha}) are suppressed by 
an additional factor $\alpha_s/\pi$. Their inclusion gives rise to 
the summation of ${\cal O} (\alpha_s\alpha^n \ln^n M_Z^2)$ effects, but 
the integration
cannot be performed analytically. Numerically this summation affects the result
at the $10^{-5}$ level which can safely be neglected.}. The shift, $-0.00030$, 
between Eqs.~(\ref{sew4}) and~(\ref{sew5}) is somewhat larger than the shift, 
$-0.00022$, obtained in Ref.~\cite{Marciano:vm}, which is in part due to 
the summation, but mainly due to the inputs. The uncertainty in 
Eq.~(\ref{sew5}) is from the current uncertainty in 
$\alpha_s = 0.120 \pm 0.002$, while other 
parametric uncertainties are miniscule\footnote{What enters Eqs.~(\ref{sew4}) 
and ~(\ref{sew5}) is the \ms $b$-quark definition, which is free of renormalon
ambiguities and therefore much better known (see Ref.~\cite{Erler:2002bu} for 
a recent sub-percent determination) than the $b$-quark pole mass. Higher order
matching corrections are also smaller if one uses the \ms mass definition.}. 
Neglecting two-loop ${\cal O} (\alpha^{n+1} \ln^n M_Z^2)$ effects,
Eq.~(\ref{sew5}) simplifies,
\be
\ba{c}
   S(m_\tau,M_Z) =
   \left[{\alpha(m_b)\over\alpha(m_\tau)}\right]^{{ 9\over 19}}
   \left[{\alpha(M_W)\over\alpha(m_b)}   \right]^{{ 9\over 20}}
   \left[{\alpha(M_Z)\over\alpha(M_W)}   \right]^{{36\over 17}}
\vspace*{8pt} \\  
   \left[{\alpha_s(m_b)\over\alpha_s(m_\tau)}\right]^{{3\over 25}
   {\alpha(m_\tau)\over\pi}}
   \left[{\alpha_s(M_Z)\over\alpha_s(m_b)}   \right]^{{3\over 23}
   {\alpha(m_b)\over\pi}},
\label{sew5b}
\ea
\ee
which differs by only $\approx 3\times 10^{-6}$.  Neglecting further
the numerically similar three-loop ${\cal O} (\alpha_s\alpha^n \ln^n M_Z^2)$
effects, one can expand the second line in Eq.~(\ref{sew5b}) to linear 
order in $\alpha$.  If one then rewrites the resulting expression in terms
of QCD scale parameters, $\Lambda_{\rm QCD}$, one encounters the 
double-logarithmic form originally obtained 20 years ago~\cite{Sirlin:1981ie}.

To summarize, next-to-leading order effects {\em reduce\/} the leading order
summation by about 50\%, {\it i.e.}, they are numerically of the same order.
Both effects are in turn numerically of order $\alpha$, so they must be included
for a complete ${\cal O} (\alpha)$ evaluation.  Unknown higher orders are 
suppressed by at least a factor of $\alpha_s/\pi$ relative to any of the effects 
mentioned before.  Thus, the uncertainty due to higher order effects is of order 
${\cal O} (\alpha\alpha_s) \sim 0.0003$.

Following Ref.~\cite{Davier:2002dy} one can write,
\be
   S_{\rm EW} \equiv S(m_\tau,M_Z) 
                {S_{\rm EW}^{\rm sub,had} \over S_{\rm EW}^{\rm sub,lep}}
              = 1.0201 \pm 0.0003,
\label{sew}
\ee
but there $S_{\rm EW} = 1.0267 \pm 0.0027$ is quoted instead. 
Almost 2/3 of the difference is due to 
the error pointed out after Eq.~(\ref{sew2b}), and about 5\% due to neglecting
next-to-leading order contributions to $S(m_\tau,M_Z)$. Another 15\% difference
is likely due to applying Eq.~(\ref{sew4}) incorrectly (as discussed above).
The remaining 15\% can perhaps be traced to use of the on-shell definition of 
$\alpha$ in place of the \ms definition as used in the present work. Note, that
the derivation of Eq.~(\ref{sew4}) assumes a {\em mass-independent\/} 
renormalization scheme (such as the \ms scheme), in which at each fermion threshold 
the $\beta$-function coefficients change by a finite amount: this is the origin
of the product form of Eq.~(\ref{sew4}). Thus, the solution~(\ref{sew4}) cannot
be applied to {\em mass-dependent\/} schemes, such as the on-shell 
renormalization scheme. It is emphasized again, that the numerical difference 
to Ref.~\cite{Davier:2002dy} should not be viewed as a scheme-dependance and 
thus as an estimate of uncalculated higher order corrections
(which are much smaller as discussed above). On the contrary,
one should expect that a self-consistent treatment within the on-shell scheme 
will reproduce the result of the present work.  

As far as the $\tau$-based analysis of Ref.~\cite{Davier:2002dy} is concerned,
about 77\% of the data is affected by $S_{\rm EW}$.  Since $S_{\rm EW}$ 
obtained in this paper differs by about 0.65\% from the one in 
Ref.~\cite{Davier:2002dy}, one expects a 0.5\% shift in the extracted
$\Delta a_\mu^{\rm had,(2)}$.  Including an update of the CKM matrix element 
$|V_{ud}|$ entering the analysis (the value, $|V_{ud}| = 0.9752 \pm 0.0007$, 
is replaced by, $|V_{ud}| = 0.97485 \pm 0.00046$, from the fit result of 
Ref.~\cite{Hagiwara:pw}), this amounts to about one half of the current
experimental uncertainty of 0.8 parts per billion~\cite{Bennett:2002jb}
for the muon magnetic moment.
The $\tau$-based Standard Model prediction would then be consistent with 
the measurement~\cite{Bennett:2002jb} within about one standard deviation. 
The discrepancy to the $e^+ e^-$ based analysis of Ref.~\cite{Davier:2002dy} 
would correspondingly be larger. Furthermore, the smaller
errors in Eq.~(\ref{sew}) and in $|V_{ud}|$
compared to Ref.~\cite{Davier:2002dy} should lead to a slight 
reduction of the overall uncertainty of the $\tau$-based result. As a final
remark, the recent determination~\cite{Erler:2002bu} of $\alpha_s$ from 
the $\tau$ lifetime, when updated with the present next-to-leading order 
analysis, increases $\alpha_s (M_Z)$ by less than $0.0001$. 

\vspace*{33pt}

\centerline{\bf Acknowledgements:}
It is a pleasure to thank Eric Braaten and Michael Ramsey-Musolf for their
comments.


\begin{thebibliography}{99}

\bibitem{Davier:2002dy}
M.~Davier, S.~Eidelman, A.~H\"ocker and Z.~Zhang, {\em Confronting Spectral 
Functions from $e+ e-$ Annihilation and $\tau$ Decays: Consequences for the 
Muon Magnetic Moment}, e-print {\tt hep-ph/0208177}.

\bibitem{Bennett:2002jb}
Muon g-2 Collaboration: G.W.~Bennett \etal, \pru{89}{101804}(2002).

\bibitem{Sirlin:1977sv}
A.~Sirlin, \rmq{50}{573}(1978).

\bibitem{Marciano:vm}
W.J.~Marciano and A.~Sirlin, \pru{61}{1815}(1988).

\bibitem{Abbaneo:2001ix}
ALEPH, DELPHI, L3, and OPAL Collaborations, LEP Electroweak Working Group and 
SLD Heavy Flavor and Electroweak Groups: D.~Abbaneo \etal,
{\em A Combination of Preliminary Electroweak Measurements and Constraints on 
the Standard Model}, e-print {\tt hep-ex/0112021}.

\bibitem{Erler:1998sy}
J.~Erler, \prt{59}{054008}(1999).

\bibitem{Hagiwara:pw}
Particle Data Group: K.~Hagiwara \etal, \prt{66}{010001}(2002).

\bibitem{Braaten:1990ef}
E.~Braaten and C.S.~Li, \prt{42}{3888}(1990).

\bibitem{Sirlin:1981ie}
A.~Sirlin, \npb{196}{83}(1982).

\bibitem{Marciano:pd}
W.J.~Marciano and A.~Sirlin, \pru{56}{22}(1986).

\bibitem{Erler:2002bu}
J.~Erler and M.~Luo, {\em Precision Determination of Heavy Quark Masses and 
the Strong Coupling Constant}, e-print {\tt hep-ph/0207114}.

\end{thebibliography}
\end{document}